\documentclass[12pt,a4paper]{article}
\usepackage{t1enc}
\usepackage[latin1]{inputenc}
\usepackage[english]{babel}
\usepackage{graphics}
\begin{document}

\title{INVESTIGATION OF THE KERR-LENS MODE LOCKING ABILITY OF Cr:ZnSe
SOLID-STATE LASER}
\author{V.L. Kalashnikov, I.G. Poloyko}
\maketitle
\begin{center}
International Laser Center, 65 Skorina Ave., Bldg. 17, Minsk, 220027 Belarus,
tel./fax: (375-0172) 326-286, e-mail: vkal@ilc.unibel.by,
http://www.geocities.com/optomaplev
\end{center}

\begin{abstract}
The theoretical calculation for nonlinear refractive index \textit{n$_{2}$}
in Cr$^{2+}$: ZnSe - active medium predicts the strong defocusing cascaded
second-order nonlinearity within 2 - 3 $\mu$m spectral range. On the basis
of this result the optimal cavity configuration for Kerr-lens mode locking
is proposed that allows to achieve a sub-100 fs pulse duration. The
numerical simulations testify about strong destabilizing processes in the
laser resulting from a strong self-phase modulation. The stabilization of
the ultrashort pulse generation is possible due to spectral filtering that
increases the pulse duration up to 300 fs.
\end{abstract}

\section{Introduction}
The intensive investigations of the Cr$^{2+}$-doped laser media testify
about their high potential in the tunable generation in mid-IR between 2 --
5 $\mu$m. A special interest is connected with Cr$^{2+}$: ZnSe -- active
medium due to its excellent material properties: high thermal conductivity
which is close to that one of YAG, the absence of excited state absorption,
and very broad gain band which allows for tunable generation in 2 -- 3 $\mu$m 
range. The latter is very attractive for a lot of applications: remote
sensing, spectroscopy, ophthalmology, and neurosurgery.

At this moment, the pulsed$^{1}$ and continuous-wave operation$^{2}$ in Cr$%
^{2+}$: ZnSe lasers was demonstrated with InGaAs and Tm:YALO laser pumping,
which demonstrated high efficiency and favorable lasing abilities of this
active crystal.

A very interesting property of Cr$^{2+}$: ZnSe is the high coefficient of
the nonlinear refraction$^{3}$ that results from the relatively small
bandgap, i. e., in fact, from the semiconductor nature of the material. As
it is known, the higher nonlinear refraction is favorable to self-starting
of Kerr-lens mode locking$^{4}$. The decrease of Kerr-lens mode locking
threshold allows to simplify the laser configuration, to improve the
ultrashort pulse stability, and to increase the lasing efficiency. Another
advantage of Cr$^{2+}$: ZnSe as media for ultrashort pulse laser is a very
broad gain band, which can support the generation of 12 fs pulses.

However, high nonlinear refraction coefficient is the source of some
shortcomings. As it was shown$^{5}$, the Kerr-lens mode-locked lasers
demonstrate a variety of unstable regimes that can destroy the ultrashort
pulse and prevent from the pulse shortening.

Here, for first time to our knowledge we analyze the Kerr-lens mode locking
abilities of Cr$^{2+}$: ZnSe laser. The analysis of ZnSe nonlinear
properties is presented that is necessary for an optimization of the laser
design and for investigation of self-starting ability and ultrashort pulse
stability in Cr$^{2+}$: ZnSe laser.

\section{Nonlinear refraction in an active medium}
As it is known$^{3}$, the two-photon absorption, Raman scattering and Stark
effect strongly contribute to nonlinear refractivity in semiconductors. The
main parameter, which defines a value of nonlinear refraction coefficient $%
n_{2}$, is the bandgap energy $E_{g}$:

\[
n_{2}\,[esu]=\frac{K\sqrt{E_{p}}G_{2}(\frac{\hbar \omega }{E_{g}})}{%
n_{0}E_{g}^{4}}.
\]

Here $n_{0}$ is the linear coefficient of refraction; $E_{p}=21\;eV$ for the
most of direct gap semiconductors; $K=1.5\cdot 10^{-8}$ is a
material-independent constant; $G_{2}$ is the function depending only on the
ratio of photon energy and the energy gap of the material. For ZnSe $%
E_{g}=2.58\;eV$. For $G_{2}$ we used the following approximation in the below
bandgap region:

\[
G_{2}(x)=0.01373+\frac{0.656\cdot 10^{-14}}{x^{2}}+\frac{0.1889\cdot 10^{-37}%
}{x^{6}},
\]%
which was obtained from experimental data presented in Ref$^{4}$. In
addition to dependence of $G_{2}$ on wavelength, we have to take into
account the dispersion of $n_{0}$. For ZnSe the next approximation is valid:

\[
n_{0}=2.4215+\frac{0.4995\cdot 10^{-7}}{\lambda }-\frac{0.1747\cdot 10^{-13}%
}{\lambda ^{2}}+\frac{0.3429\cdot 10^{-19}}{\lambda ^{3}}
\]%
(here $\lambda $ is in meters). A resulting dependence of $n_{2}$ on
wavelength is shown in Fig. 1. There is a strong Kerr nonlinearity producing
self-focusing effect. The value of $n_{2}$ exceeds the typical values of the
most of dielectric crystals by two orders of magnitude.

However, this is not sufficient for the estimation of the nonlinear
properties of ZnSe. Despite the fact that ZnSe has cubical lattice, there is
not center of inversion in crystal. As result, the medium possesses the
second-order nonlinear properties. The corresponding nonlinear coefficient $%
d=80\;pm/V\;^{6}$. Although there is no efficient frequency conversion, the
existence of second-order nonlinearity can strongly contributes to the
nonlinear refractivity due to so-called cascaded second-order nonlinearity,
which can be used for Kerr-lens mode locking$^{7}$. Usually, change of the
index of refractivity has non-Kerr character, but the Kerr-like
approximation $n=n_{0}+n_{2}I$ (where $I$ is the field intensity) is valid if%
$^{6}$ $\frac{\left( \frac{\Delta k}{2\chi }\right) ^{2}}{I}<<1$ , where

$\chi =\frac{2\omega d}{\sqrt{2\epsilon _{0}n_{0}^{2}n_{2\omega }c^{3}}}$ ; $%
\Delta k=2k_{0}-k_{2\omega }$; $\omega $, $k_{0}$ are the fundamental
frequency and wave number, correspondingly; $n_{2\omega }$, $k_{2\omega }$ are
the linear refractive index and wave number for the second harmonic. Our
calculations showed, that the Kerr approximation for ZnSe is valid up to
intensities of $2.2\;TW/cm^{2}$ at 2 $\mu$m and $600\;GW/cm^{2}$ at 3 $\mu$m, 
and it is used throughout present work.

The nonlinear coefficient of refraction due to cascading second-order
nonlinearity is$^{6}$: 

\[
n_{2}=\frac{4\pi d^{2}}{\epsilon _{0}n_{0}^{2}n_{2\omega }c\lambda \Delta k}.
\]

Note, that in the case of the normal dispersion of $n_{0}$ the nonlinearity
due to second-order nonlinearity has a defocusing nature. The dependence of $%
n_{2}$ on the wavelength is shown in Fig. 2. One can see, that the nonlinear
refraction is very strong. As result of the joint contribution of third- and
second-order nonlinearities, a net-coefficient of nonlinear refraction is
shown in Fig. 3. The nonlinearity is focusing up to 1.2 $\mu$m and
defocusing in mid-IR range. An experimental value of $n_{2}$ measured at
1.06 $\mu$m demonstrates an excellent agreement with the corresponding
value of calculated curve in Fig. 3.

\section{Laser configuration}
The obtained value of $n_{2}$ enables us to estimate a mode locking
efficiency due to Kerr-lensing. We have used the method presented in $^{8}$:
for each fixed value of the focal length of the mirror $M_{2}$ (see Fig. 4)
we varied a folding distance and the position of active medium trying to
maximize the value of $\frac{\partial T}{\partial I}$ , where $T$ is the
effective transmission of the hard aperture. This parameter defines the
inverse saturation intensity of the efficient fast saturable absorber,
induced by joint action of Kerr self-focusing in active medium and
difractional losses at aperture. For the common Kerr-lens mode-locked lasers 
$\frac{\partial T}{\partial I}\approx 10^{-10}\,-\,10^{-12}\,cm^{2}/W$ . In
the case of ZnSe the modulation parameter is shown in Fig. 5. It is seen,
that the Kerr-lensing efficiency in this case is higher by two orders of
magnitude. It is interestingly, that the defocusing nature of the
nonlinearity does not affect the process of the cavity optimization
essentially: the change of the sign of $n_{2}$ only slightly affects the
position of active medium and folding distance. The optimized cavity
configuration is shown in Fig. 4.

\section{Ultrashort pulse stability}
Calculated values of $\frac{\partial T}{\partial I}$ suggest a good
Kerr-lens mode locking ability of Cr$^{2+}$: ZnSe -- laser. However, the
strong self-phase modulation while being the positive factor facilitating
self-focusing, can, on the other hand, destabilize an ultrashort pulse
generation due to automodulational instability$^{5}$. In order to estimate
the mode locking stability in Cr$^{2+}$: ZnSe -- laser we performed the
numerical simulation on the basis of fluctuation model and distributed
scheme of the laser generation. The master equation was:

\[
\frac{\partial a}{\partial z}=\left( \alpha +(t_{f}+iD)\frac{\partial ^{2}}{%
\partial t^{2}}-i\beta \left| a\right| ^{2}-\frac{\gamma }{1+\sigma \left|
a\right| ^{2}}\right) a,
\]

where $a$ is the field, $z$ is the longitudinal coordinate, $\alpha$ is the
saturated by full field energy gain coefficient, $t_{f}$ is the inverse
bandwidth of the spectral filter, $D$ is the net-group delay dispersion, $%
\beta =\frac{2\pi n_{2}z_{g}}{\lambda n_{0}}$ is the self-phase modulation
parameter, $z_{g}$ is the active medium length, $\gamma $ is the diffraction
loss, defined from configuration of laser, $\sigma $ = $\frac{\partial T}{%
\partial I}$ .

The strong self-phase modulation and self-focusing dominate over the gain
saturation. The last factor provides the negative feedback in laser that
stabilizes ultrashort pulse. The critical parameter defining the ultrashort
pulse dynamics is $\tau =\frac{t_{f}}{\beta E_{s}}$ , where $E_{s}$ is the
gain saturation fluency. The decrease of this parameter worsens the pulse
stability. To comparison, in ZnSe $\tau $ is ten times lower than in Ti:
sapphire. Therefore, to obtain the stable pulse generation, we have to
increase $\tau $ e.g. by increasing $t_{f}$. The stable pulses were obtained
for $t_{f}=120\;fs$. The pulse duration was about 300 fs for 800 mW absorbed
pump power. An additional stabilization factor is the minimization (down to
zero) of net-dispersion.

The character of the pulse destabilization needs some comments. The typical
scenarios are shown in Figs. 6 -- 8. The pulse behaviors presented here were
calculated for the same set of the laser parameters but for different
initial noise samples. There are the possibilities for multipulse generation
(Fig. 6), generation of the breezer-like pulse and its subsequent
disintegration (Fig. 7), and the collapse of the pulse into several pulses
(Fig. 8). Note, that breezer-like pulse is formed without action of the
soliton mechanism in laser because of net-dispersion in this case is close
to $0$. The regime depicted in Fig. 8 is characterized by low-frequency
modulation of the generation spectrum that at last leads to the pulse
collapse. It should be noted, that the generation of the stable single pulse
is depended on the random noise sample, too.

\section{Conclusion}
The analysis of the nonlinear properties of ZnSe -- crystal was performed.
It was found, that the strong second-order cascading nonlinearity in the
combination with third-order phase nonlinearity causes a strong focusing
Kerr nonlinearity in near-IR and defocusing quasi-Kerr nonlinearity in
mid-IR ranges. As result, the efficiency of the Kerr-lensing in active
medium is two-three orders of magnitude higher than in typical active media
of femtosecond lasers. The optimal cavity configuration for Kerr-lens mode
locking was found. But, as the analysis testifies, the main problem of the
ultrashort pulse generation in Cr$^{2+}$: ZnSe -- laser is the pulse
destabilization due to strong self-phase modulation. This requires to
decrease the bandwidth of spectral filter. The last allows to generate 300
fs pulses. The self-start of Kerr-lens mode locking has a statistical
character, that is depended on the noise sample. The investigation of the
pulse destabilization testifies about formation of the breezer-like pulses
and pulse collapse due to low frequency perturbations.

\section{References}

1. R.H. Page, J.A. Skidmore, K.I. Schaffers et al., Trends in Optics and Photonics, (OSA, Washington, 1997).\\
2. G. J. Wagner, T. J. Carring, R. H. Page et al., Opt. Lett., 24, 19 (1999).\\
3. M. Sheik-Bahae, D. C. Hutchings, D. J. Hagan, and E. W. van Stryland, IEEE J. Quantum Electr., 27, 1296 (1991).\\
4. C. Radzewicz, G. W. Pearson, and J. S. Krasinski, Optics. Commun., 102, 464 (1993).\\
5. J. Jasapara, V. L. Kalashnikov, et al., J. Opt. Soc. Am., B 17, 319 (2000).\\
6. R. L. Sutherland, Handbook of nonlinear optics (NY, 1996).\\
7. G. Cerullo, S. De Silvestry, A. Monguzzi et al., Opt. Letts., 20, 746 (1995).\\
8. V. L. Kalashnikov, V. P. Kalosha et al., J. Opt. Soc. Am., B 14, 964 (1997).

\clearpage

\begin{figure}
	\begin{center}
		\includegraphics{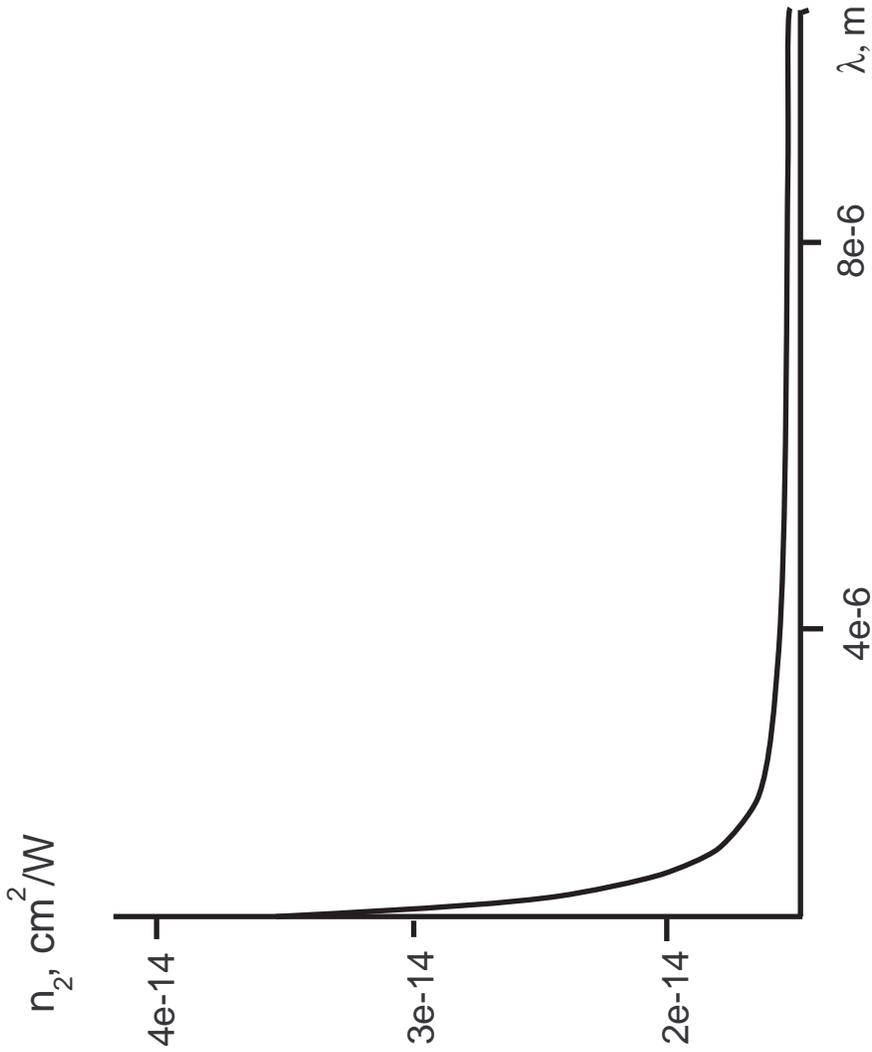}
	\end{center}

	\caption{The dependence of n$_{2}$ on wavelength in the case of third-order nonlinearity contribution}
\end{figure}

\begin{figure}
	\begin{center}
		\includegraphics{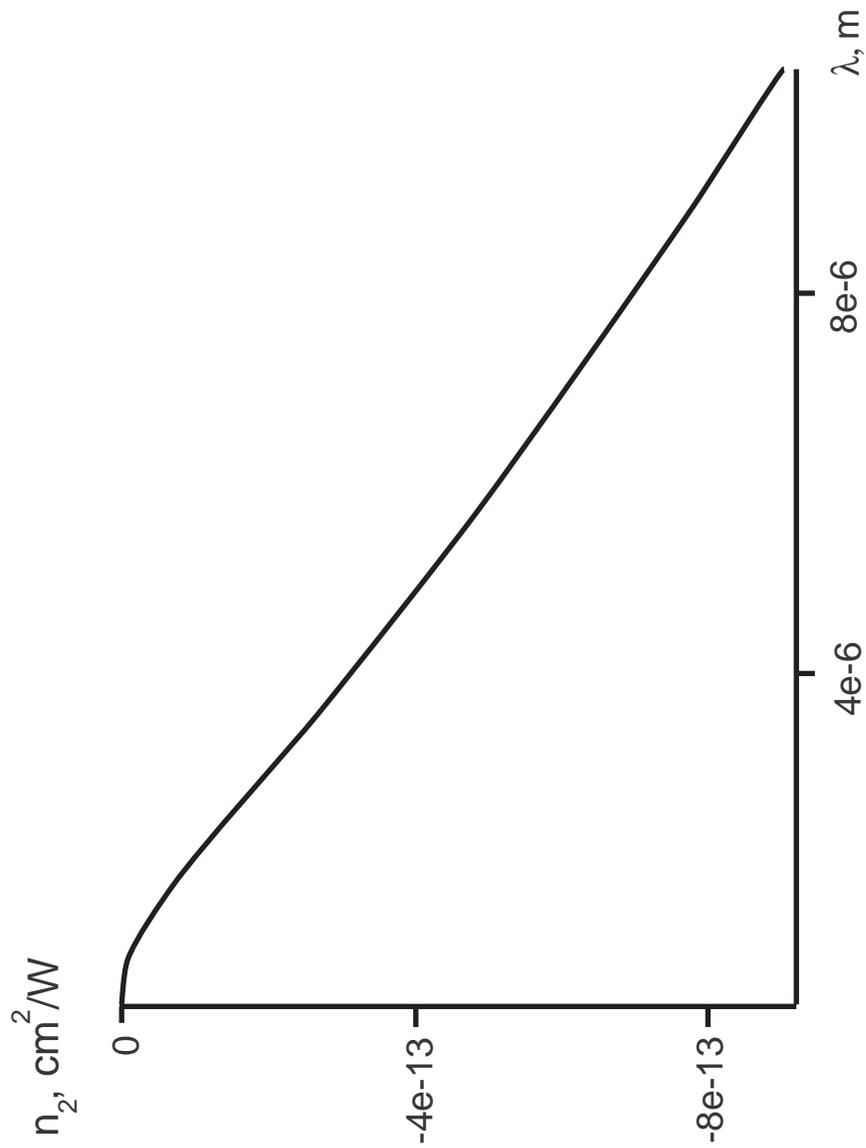}
	\end{center}

	\caption{The dependence of n$_{2}$ on wavelength in the case of second-order nonlinearity contribution}
\end{figure}

\begin{figure}
	\begin{center}
		\includegraphics{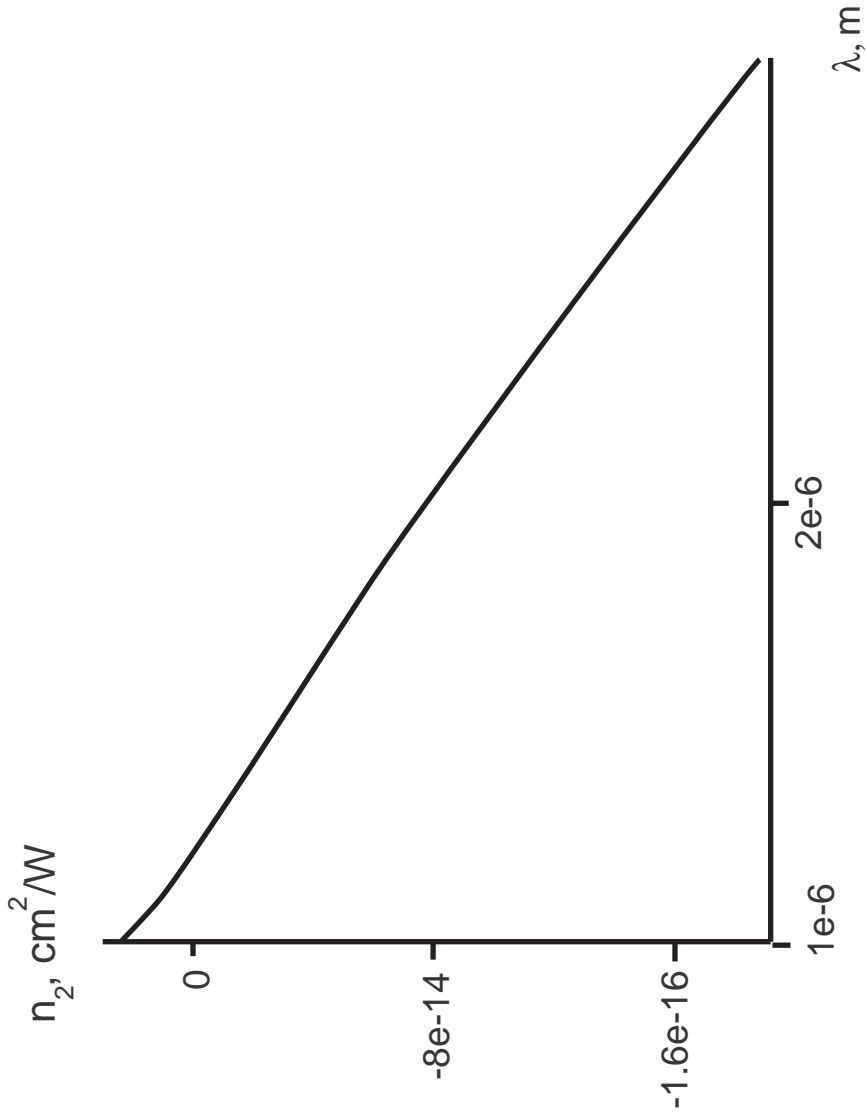}
	\end{center}

	\caption{Net-coefficient of nonlinear refraction in ZnSe}
\end{figure}

\begin{figure}
	\begin{center}
		\includegraphics{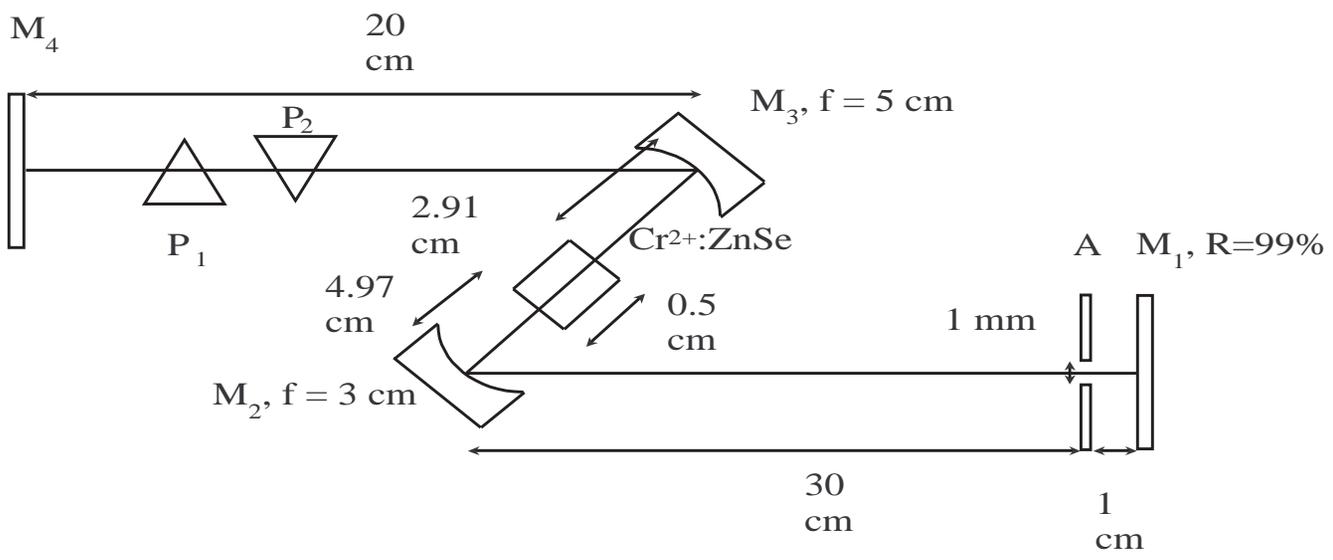}
	\end{center}

	\caption{Optimal cavity design for Kerr-lens mode locking}
\end{figure}

\begin{figure}
	\begin{center}
		\includegraphics{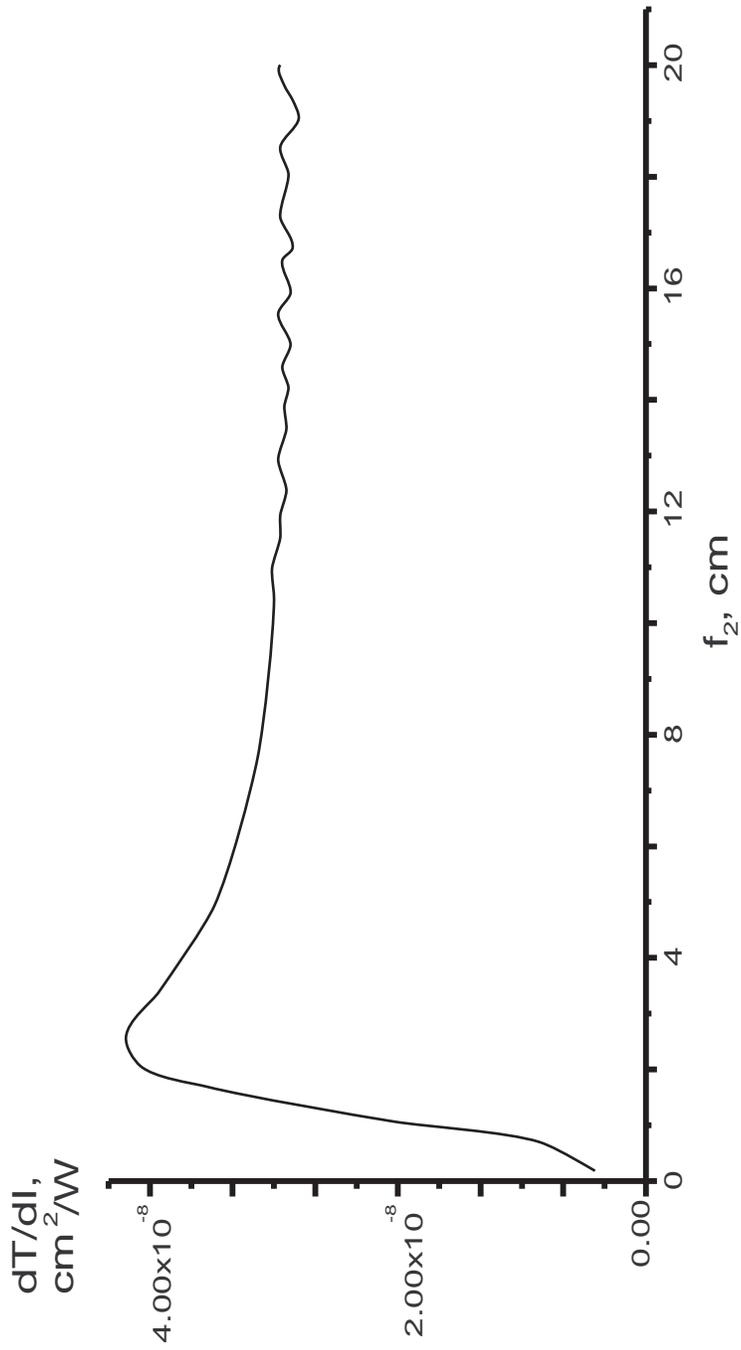}
	\end{center}

	\caption{Kerr-lens mode locking efficiency in ZnSe - laser}
\end{figure}

\begin{figure}
	\begin{center}
		\includegraphics{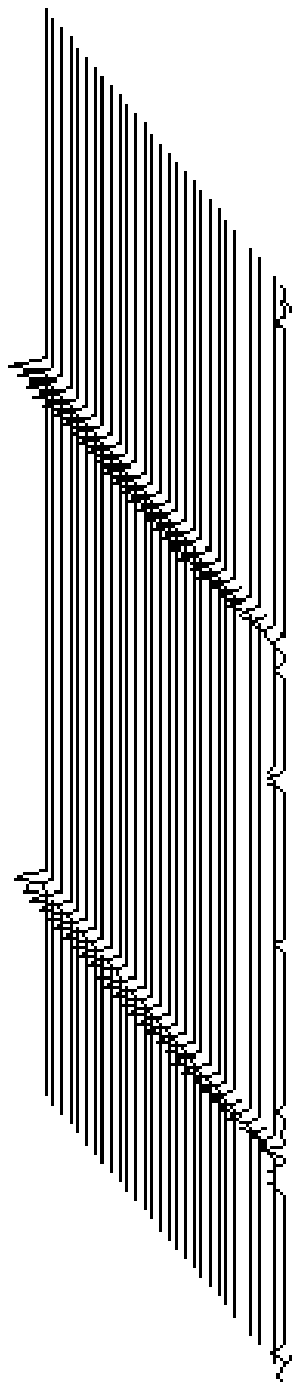}
	\end{center}

	\caption{Pulse train}
\end{figure}

\begin{figure}
	\begin{center}
		\includegraphics{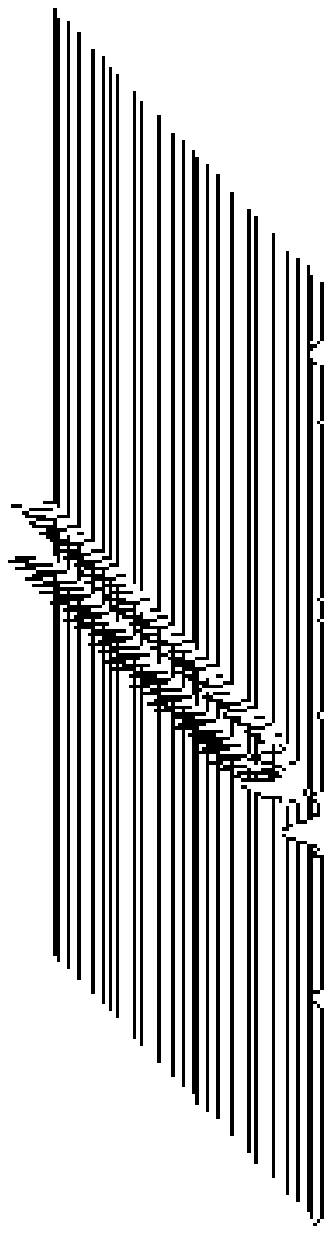}
	\end{center}

	\caption{Pulse train}
\end{figure}

\begin{figure}
	\begin{center}
		\includegraphics{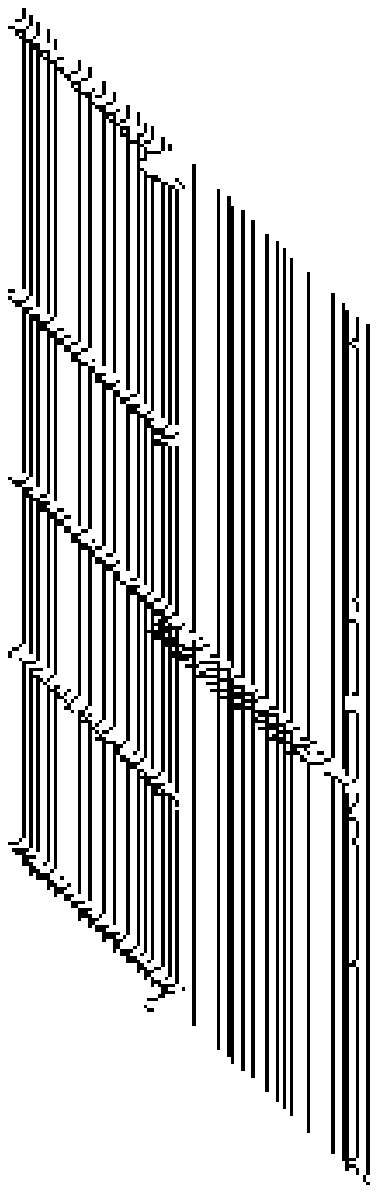}
	\end{center}

	\caption{Pulse train}
\end{figure}

\end{document}